# The Interplay of Covalency, Hydrogen Bonding and Dispersion Leads to a Long Range Chiral Network: The Example of 2-Butanol

Melissa L. Liriano,[†] Javier Carrasco,[‡] Emily A. Lewis,[†] Colin J. Murphy,[†] Timothy J. Lawton,[†] Matthew D. Marcinkowski,[†] Andrew J. Therrien,[†] Angelos Michaelides,[∥] E. Charles H. Sykes[*,†]

[†] Department of Chemistry, Tufts University, Medford, Massachusetts 02155, USA
[‡] CIC Energigune, Albert Einstein 48, 01510 Miñano (Álava), Spain
[∥] Thomas Young Centre, London Centre for Nanotechnology and Department of Chemistry, University College London, London WC1E 6BT, United Kingdom
[*] Corresponding author: charles.sykes@tufts.edu

**ABSTRACT**

The assembly of complex structures in nature is driven by an interplay between several intermolecular interactions, from strong covalent bonds to weaker dispersion forces. Understanding and ultimately controlling the self-assembly of materials requires extensive study of how these forces drive local nanoscale interactions and how larger structures evolve. Surface-based self-assembly is particularly amenable to modeling and measuring these interactions in well-defined systems. This study focuses on 2-butanol, the simplest aliphatic chiral alcohol. 2-butanol has recently been shown to have interesting properties as a chiral modifier of surface chemistry, however, its mode of action is not fully understood and a microscopic understanding of the role non-covalent interactions play in its adsorption and assembly on surfaces is lacking. In order to probe its surface properties we employed high-resolution scanning tunneling microscopy and density functional theory simulations. We found a surprisingly rich degree of enantiospecific adsorption, association, chiral cluster growth and ultimately long range, highly ordered chiral templating. Firstly, the chiral molecules acquire a second chiral center when adsorbed to the surface via dative bonding of one of the oxygen atom lone pairs. This interaction is controlled via the molecule's intrinsic chiral center leading to monomers of like chirality, at both chiral centers, adsorbed on the surface. The monomers then associate into tetramers via a cyclical network of hydrogen bonds with an opposite chirality at the oxygen atom. The evolution of these square units is surprising given that the underlying surface has a hexagonal symmetry. Our DFT calculations, however, reveal that the tetramers are stable entities that are able to associate with each other by weaker van der Waals interactions and tessellate in an extended square network. This network of homochiral square pores grows to cover the whole Au(111) surface. Our data reveals that the chirality of a simple alcohol can be transferred to its surface binding geometry, drive the directionality of hydrogen bonded networks and ultimately extended structure. Furthermore, this study provides the first microscopic insight into the surface properties of this important chiral modifier and provides a well-defined system for studying the network's enantioselective interaction with other molecules.

## 1. INTRODUCTION

Although hydrogen bonding and van der Waals interactions are relatively weak when compared to covalent bonds, it has been found that a careful balance between all these forces can lead to a rich evolution of structures over many length scales. For example, they allow stable and highly ordered 2-dimensional self-assembled monolayers (SAMs) to form on surfaces [1]; are responsible for the cohesion properties of several crystalline phases of ice [2]; and drive the folding of proteins and nucleic acids into their functional forms.[1] The interplay between these intermolecular forces is, therefore, relevant to the fields of physics, biology, and chemistry. There is an increasing interest in using SAMs as building blocks for making functional nanomaterials, hence, understanding how to manipulate the delicate balance between these noncovalent interactions is of great interest to the fields of material science, microelectronics and nanotechnology.[1] In terms of hydrogen bonding, water is by far the most studied and perhaps the best understood molecule in terms of its bulk and surface-bound structures.[3-17] Methanol and larger alcohols represent simple systems for understanding hydrogen-bonded networks between molecules with isolated hydroxyl groups.[18-21] We have previously shown that the adsorption of methanol and asymmetric thioethers on surfaces leads to the emergence of chirality in the surface-bound molecules.[18-21, 22-25] The chiral center in adsorbed methanol and asymmetric thioethers arises from dative binding of one of the two lone pairs on the O or S atom, respectively, to the surface resulting in the adsorbed molecule having four different groups around a central atom and hence chirality. Interestingly, surface-bound enantiomers adsorbed in this manner interact enantiospecifically with each other and their surface-mediated self-assembly results in a rich array of both highly ordered homochiral and racemic structures.[18-21, 22-25] In this study, we extend these ideas and explore the adsorption and assembly of an intrinsically chiral alcohol.

Fundamentally, 2-butanol (2-BuOH) represents the simplest chiral alcohol and from a practical standpoint it has been shown to interact enantioselectively with catalytically relevant chiral modifiers.[26-27] For example, temperature programmed desorption (TPD) was used to study L-Proline/Pd and α-(1-Naphthyl)ethylamine (NEA)/Pd surfaces when using R- and S-2-BuOH as probes for surface chirality.[26-27] Interestingly, it was found that on the L-Proline/Pd surface, 2-BuOH was regenerated enantioselectively from 2-butoxide decomposition products and pre-adsorbed deuterium.[26] On the NEA/Pd modified surface, hydrogen bonding between the hydroxyl group of the chiral alcohol and the amine of NEA enabled enantioselective chemisorption of 2-BuOH.[27] Furthermore, 2-BuOH adsorption has been shown to bestow enantioselective properties on certain surfaces.[26-32] TPD was used to probe enantioselective properties of R- and S-2-BuOH adlayers when 2-BuOH was used as a chiral modifier on Pt, Pd, and Au/Pd surfaces.[28-32] Initially it was hypothesized that the controlled decomposition of small amounts of 2-BuOH to 2-butoxide on Pt(111) and Pd(111) surfaces facilitated enantiospecific adsorption of chiral probe molecules.[28-29] It was proposed that at low coverages, 2-butoxide formed chiral "pockets" capable of "chiral docking" incoming reactants through steric hindrance.[28] Further experimentation, however, revealed that lower amounts of 2-butoxide and specific concentrations of 2-BuOH adsorbed on the surface lead to increased enantioselective chemisorption of chiral reactants.[30] It was proposed that hydrogen bonding interactions between the labile proton in 2-BuOH with the incoming probe molecule drives enantiospecific adsorption.[30] In order to gain a more complete understanding of these important enantiospecific interactions of 2-BuOH on surfaces a molecular-scale picture of how it adsorbs and assembles is required, but has not been achieved to date.

Towards this end, low-temperature scanning tunneling microscopy (STM) along with density functional theory (DFT) calculations and STM image simulation was used to examine the adsorption and assembly of R- and S-2-BuOH on a Au(111) surface. Au(111) was chosen as the first surface of interest as it is the least reactive metal and thus provides an opportunity to understand the adsorption and assembly of the intact chiral alcohol. Surprisingly, unlike methanol and water which form hexamers on six-fold symmetric surfaces, we found that the predominant arrangement of 2-BuOH are tetramers, consisting of four 2-BuOH molecules of like surface-bound chirality hydrogen-bonded together. Similar to methanol and water clusters, the hydrogen bonding in 2-BuOH tetramers is directional and induces a rotation of the chiral clusters with respect to the underlying substrate. Van der Waals (vdW) interactions between the hydrogen bonded tetramers drive the formation of highly ordered square networks.

The remainder of the paper involves a description of the experimental and computational methods employed (section 2). Following this, in section 3, we present high-resolution images of 2- BuOH overlayers, a determination of the key molecular scale building block within the overlayers, and then we discuss the large scale assembly of the overlayers. We close in section 4 with some conclusions and a brief perspective for future work.

## 2. EXPERIMENTAL SECTION

**STM Experiments.** All STM experiments were performed with an Omicron NanoTechnology low temperature (LT) STM. The base pressure in the STM chamber was 1 x $10^{-11}$ mbar. The Au(111) single crystal sample was cleaned by cycles of $Ar^+$ sputtering (14 μA, 1 kV) and annealing (1000 K). The sample was then transferred into the pre-cooled STM stage with a base temperature of 5 K. Pure enantiomers of R- and S-2-BuOH (99.9%) were purchased from Sigma Aldrich and further purified through freeze-pump-thaw cycles. Submonolayer and monolayer coverages of 2-BuOH were deposited through a high-precision leak valve onto the sample held at 5 K followed by a thermal anneal to 100 K to equilibrate the molecular assemblies. The sample was then cooled back down to 5 K to acquire high-resolution images. STM images were obtained with Omicron etched W tips at bias voltages between ±50 mV and ±300 mV and tunneling currents between 10 pA and 200 pA.

**Theoretical methods.** DFT calculations were performed using a supercell approach and the optB88-vdW functional [33] as implemented in the Vienna ab initio simulation package (VASP, version 5.3.3).[34] The optB88-vdW functional is a revised version of the vdW-DF of Dion et al.[35] which has been shown to perform well for a broad range of systems, including hydrogen-bonded adsorption systems such as the one considered here.[19-20, 36-38] We replaced the inner electrons by projector augmented wave potentials,[39] whereas the valence states were expanded in plane-waves with a cut-off energy of 500 eV. We considered metal slabs cut along the unreconstructed (111) direction consisting of 3 atomic layers thickness separated by 1.5 nm of vacuum. The metal atoms in the bottom layer were fixed to the bulk optB88-vdW optimal position ($a_{Au}$ = 0.4158 nm). The adsorption of 2-BuOH monomers, dimers, tetramers, and hexamers were modeled on 6×6 supercells. A 2×2×1 Monkhorst-Pack k-point mesh was used within this 6×6 supercell. This setup is similar to what was employed for methanol adsorption on closed packed transition metal surfaces, including Au(111), and guarantees a sufficiently tight convergence in adsorption energies and equilibrium distances.[21] We applied a dipole correction along the direction perpendicular to the metal surface and geometry optimizations were performed with a residual force threshold of 0.025 eV/Å. Adsorption energies per molecule are defined with respect to the total energy of the gas phase 2-BuOH molecule and the total energy of the unreconstructed Au(111) surface. Favorable (exothermic) adsorption corresponds to negative adsorption energy.

STM images were simulated using the Tersoff-Hamann approach[40], with a voltage of 0.1 V (corresponding to filled states) at a height of 0.66 nm above the metal surface.

## 3. RESULTS AND DISCUSSION

**A. The Au(111) Reconstruction**: Before discussing our results for 2-BuOH adsorption, it is useful to briefly comment on the structure of the clean Au(111) surface. It is well known that a clean Au(111) surface reconstructs. The unit cell of the reconstructed surface consists of 23 atoms sitting on 22 bulk lattice sites, creating a long-range elastic lattice strain in the top-most atomic layer. The surface adopts a 22 x √3, so-called herringbone reconstruction, which involves a 4.5% contraction along close-packed, or [1$\bar{1}$0], directions forming stacking faults consisting of wider fcc and narrower hcp packed regions as seen in Figure 1a.[41] The fcc-hcp stacking transitions are separated by pairs of parallel corrugation lines, called soliton walls. The distance between neighboring pairs of solitons is 6.3 nm on clean Au. The surface layer Au atoms rest in a variety of sites, with fcc and hcp atoms sitting on three-fold hollow sites while soliton wall atoms sit topographically higher on "quasibridge" sites.[41] STM images of the clean Au(111) surface herringbone reconstruction shows that soliton walls appear as bright zig-zag lines that run in three equivalent directions to relieve strain isotropically, as seen in Figure 1a.

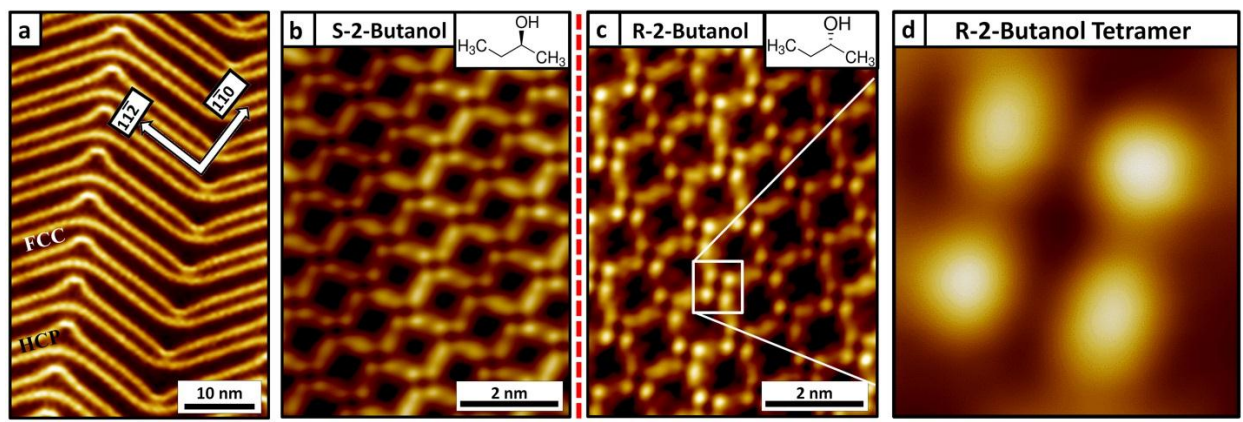

**Figure 1.** (a) High-resolution STM image of a clean Au(111) surface and its 22 x √3 herringbone reconstruction: soliton walls appear as pairs of bright zig-zag lines separating fcc and hcp stacking regions. The high symmetry [11$\bar{2}$] and [1$\bar{1}$0] directions are also highlighted. (b, c) High-resolution STM images of enantiomeric domains of S- and R-2-BuOH in the 1 ML regime. Insets show the structure of each enantiomer. The plane of symmetry (red dotted line) is aligned with the √3, or the [11$\bar{2}$], direction of the underlying Au(111) surface. The S-2-BuOH domain is rotated +25° (clockwise) and the R-2-BuOH is oriented -25° (anticlockwise) from the √3 direction as indicated by the dashed red line. (d) High-resolution image of R-2-BuOH tetramer that is highlighted by a white square in panel (c).

**B. Towards a molecular level model of 2-BuOH adsorption:** Initial experiments on 2-BuOH adsorption involved depositing S-2- BuOH on Au(111) at 5 K, annealing to 100 K to equilibrate the molecular ensembles, and imaging at 5 K for optimum stability and resolution. Figure 1(b) shows a high-resolution STM image of an overlayer that forms following this procedure. It can immediately be seen from this that highly ordered square arrays of S-2-BuOH form and that these extend over large areas of the surface. Extensive STM imaging of over 10,000 nm$^2$ of the Au(111) surface revealed that these square networks of S-2-BuOH exist in just three different rotational domains rotated 120° from each other at all surface coverages explored. Differences in the background color contrast in the STM images of the square arrays are due to the herringbone reconstruction, which leads to portions of the 2-BuOH network adsorbed over the soliton walls to appear brighter. In order to examine the chirality of these structures we performed control experiments with the opposite enantiomer, R-2-BuOH. Identical structures and rotational domains were observed when R-2-BuOH was adsorbed on the Au substrate but with opposite mirror symmetry. Figures 1 b-c are zoomed-in images of one pair of enantiomeric domains of R- and S-2-BuOH, respectively. The mirror plane of symmetry is depicted as a red dotted line in Figure 1. Measurements confirmed that the R- and S-2-BuOH domains are rotated the same angle from the Au $[11\bar{2}]$ axis but in opposite directions, with R-2-BuOH rotated -25° from the vertical √3 direction and the S-2-BuOH enantiomer rotated +25° (Figures 1b-c). Locally, the Au(111) surface has six-fold rotational symmetry and reflection mirror symmetry, however, when either R or S 2-BuOH is adsorbed one finds that the reflection symmetry planes of the Au(111) surface are destroyed due to this rotation from the high symmetry axes. Therefore, the 2-BuOH ordered domains impart a chiral footprint on the underlying surface.[42-50]

In a recent study, diffraction data was collected for a racemic sample of 2-BuOH that was crystallized under high-pressure conditions.[51] The data revealed that the 2-BuOH enantiomers phase-segregated and two levels of ordering that ultimately created interpenetrated chiral helices from homochiral chains were identified. The first level of ordering involved the assembly of homochiral helical chains consisting of 2-BuOH monomers hydrogen-bonded to adjacent molecules with the same intrinsic chirality.[51] Consistent with our previous work involving hydrogen-bonded networks on surfaces, the "tilt" or rotational angle of the hydrogen-bonded homochiral helical chains was dictated by the direction of the hydrogen–bonded network. At the second level of ordering 2-BuOH homochiral helices of the same inherent chirality interacted via van der Waals interactions to form densely packed extended structures.[51] Similarly, in this study on Au(111), examination of the high-resolution STM images of the ordered 2-BuOH domains reveals that, at the monolayer coverage, there also exists two levels of ordering. Unlike with the crystallized aggregates of 2-BuOH, however, the first level of ordering is based upon small motifs that appear in STM images as four bright protrusions.

Our previous work on methanol adsorption provides relevant insight that helps in developing a plausible adsorption model. More specifically, in methanol adsorption the O-H bonds lie almost parallel to the surface with the oxygen atoms located close to atop sites. This adsorption mode allows two hydrogen bonds between adjacent molecules and for a fairly strong interaction with the surface via an oxygen lone pair. 2-BuOH is in principle capable of forming a similar adsorption structure and so with this in mind we performed an extensive set of DFT calculations for 2-BuOH adsorption. Given that the STM images clearly identify a structure with 4 lobes we put a particular emphasis on establishing structures for adsorbed tetramers. A high-resolution STM image of a tetramer is shown in Figure 2a along with a simulated STM image (Figure 2b) and a DFT calculated structure (Figure 2d) for the most stable adsorbed tetramer. The tetrameric structure from top and side views is also shown in Figure 3c. As expected, based on our previous work on methanol adsorption, the most stable tetramer structure identified is indeed a hydrogen bonded structure with the O-H bonds close to and almost parallel to the surface. The DFT calculations also clarify the adsorption site, and as with water and methanol, we find that the 2-BuOH molecule bonds to the surface primarily through the oxygen atom with the oxygen atom preferentially adsorbing near an atop site (i.e. above a single Au atom). The DFT simulated STM image of a tetramer also indicates that the bright lobes observed in STM correspond to the raised methyl groups.

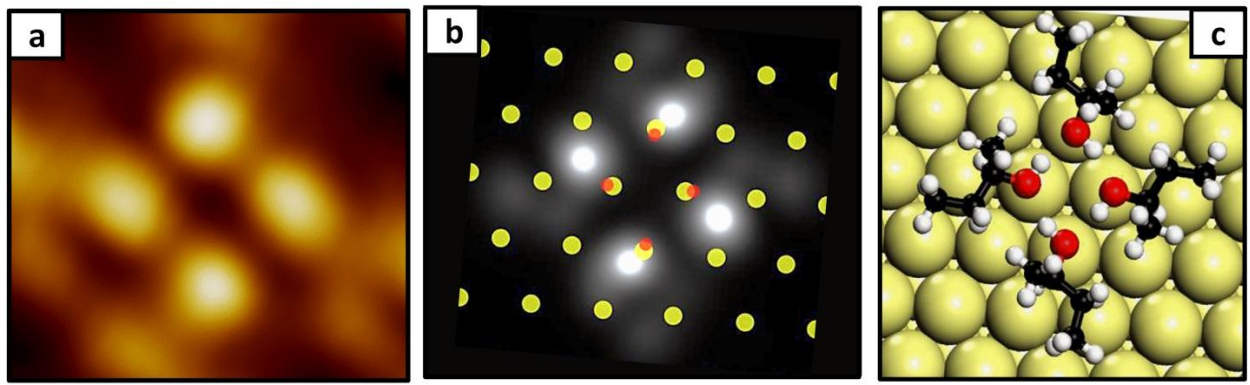

**Figure 2.** (a) High resolution STM image of an R-2-BuOH tetramer, the basic unit of the extended chiral domains. (b) Simulated STM image (at height 0.66 nm, V= 0.1 V); the yellow and red dots represent the first surface layer of Au atoms and O oxygen atoms of 2-BuOH, respectively. (c) Optimized DFT structure of an R-2-BuOH tetramer unit illustrating that each oxygen atom binds on near atop sites and that cluster formation is driven by hydrogen bonding. Color code: Au, yellow; C, black; O, red; and H, white.

Due to the symmetry of the underlying surface, one might expect the most stable cluster to be a hexamer, as was previously observed for methanol and water on (111) surfaces.[18-21] [22-25] Our DFT calculations reveal that, unlike for water and methanol in which adsorbed hexamers are clearly more stable than smaller clusters such as tetramers, for 2-BuOH on Au(111) there is essentially no difference in the stability of adsorbed hexamers and tetramers.[52] Specifically the adsorption energy of the tetramers and hexamers differs by only 5 meV/BuOH (Figure 3); a negligible difference in a DFT adsorption calculation.[53] As shown in Figure 3, however, it is clear that the tetramers and hexamers are significantly more stable than smaller clusters (monomers and dimers). This is consistent with experimental observations (as discussed below) which show that BuOH molecules aggregate into tetramers even at low coverage.

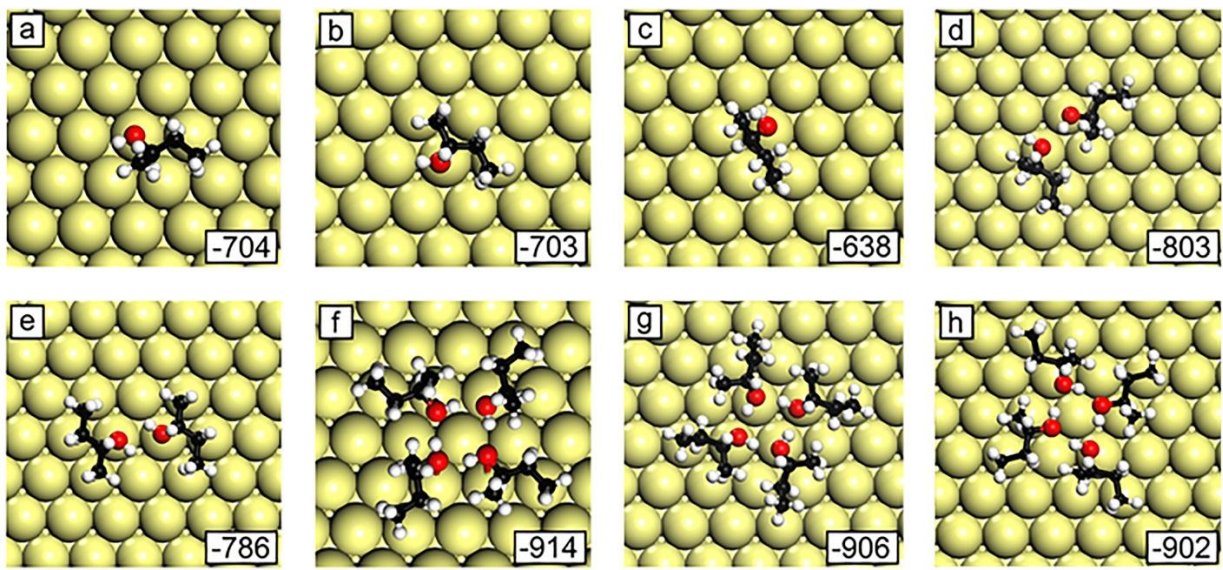

**Figure 3.** Top and side views of the most stable DFT optimized R-2-BuOH structures on Au(111): (a) monomer, (b) dimer, (c) tetramer, and (d) hexamer. The numbers indicate total adsorption energies in meV/BuOH.

From DFT calculations we can also learn something about the balance between hydrogen bonding and adsorbate substrate bonding in these systems. Interestingly, we find that the total adsorption energy of these structures is dominated by BuOH-metal interactions, but that the interaction with the surface decreases as the cluster size grows (Figure 4). In contrast, hydrogen bonding (BuOH-BuOH interactions) increases as the clusters grow, revealing a significant cooperative effect. BuOH-metal and BuOH-BuOH contributions actuate this energy in opposite directions with increasing cluster size. This balance explains why adsorption energies reach a constant value around −940 meV/BuOH after the formation of tetramers.

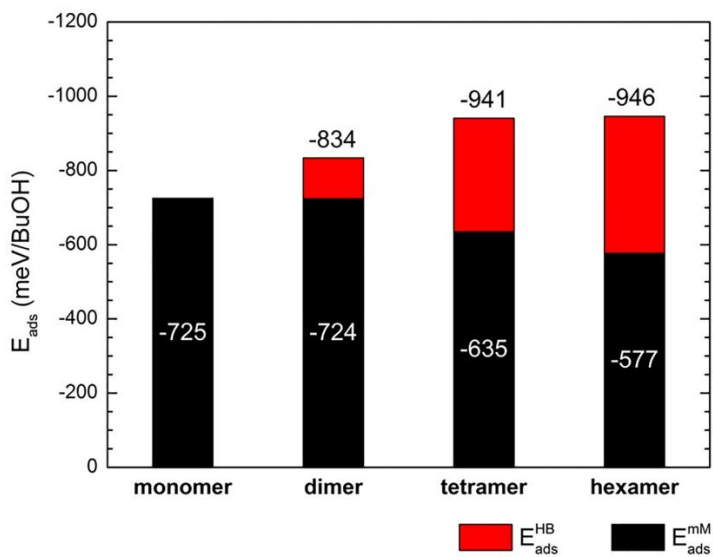

**Figure 4.** DFT computed adsorption energies ($E_{ads}$) of 2-BuOH on Au(111). BuOH-BuOH, (red) and BuOH-metal (black) contributions to the total adsorption energy of the most stable R-2-BuOH monomer, dimer, tetramer, and hexamer on Au(111). See the supporting information (SI) for details of how the energy decompositions were performed.

The 2-BuOH tetramer adopts a planar arrangement on the Au surface with each oxygen atom binding on atop Au sites and the O-H bond lies almost parallel to the surface, as illustrated in Figures 2(c) and 3(c). The proposed adsorption model indicates that there exists two levels of what is referred to in the literature as point chirality. The first level of point chirality arises when the intrinsic chirality of the molecule is preserved upon adsorption, as no molecular distortions are predicted or observed in the adsorbed state at the chiral carbon atom center ($R_c$ or $S_c$). When considering the oxygen-Au bond, another expression of point chirality arises at the oxygen atom because it is bound to a surface via a dative bond between an oxygen lone pair and the Au surface. If we consider the adsorbed oxygen atom as having four distinct groups bound to it, with the Au surface being the highest priority group and the second lone pair being the lowest priority group, then a second chiral center is formed at the adsorbed oxygen atom (Figure 5), causing each monomer to have a chirality ($R_o$ or $S_o$) associated with the oxygen atom center, in addition to point chirality at the carbon center ($R_c$ or $S_c$). This surface-bound chirality effect was previously observed for methanol hexamers imaged on Cu(111) and Au(111).[18-21]

Upon adsorption, the tetramer units are rotated slightly away from the high symmetry directions of the underlying substrate. Furthermore, the tetramers have an associated asymmetry dictated by the direction of the hydrogen-bonded network with a clockwise sense for R-2-BuOH and anticlockwise for S-2-BuOH, causing tetramers to have a surface-bound chirality. Interestingly, DFT calculations for R-2-BuOH predict that the surface-bound $R_O$-2-BuOH monomer, where $R_O$ indicates the chirality around the adsorbed oxygen atom, is 21 meV more stable than its surface-bound enantiomer, $S_O$-2-BuOH (Fig. 5, and SI). In the $S_O$-2-BuOH enantiomer, the H atom attached to C2 is pointing towards the metal surface, which imposes a geometrical strain on the bond between the O lone pair and the metal surface (Figure 5). As a consequence, for the R-2-BuOH monomer, the adsorption of $S_O$-2-BuOH is weaker than that of $R_O$-2-BuOH. While this enantioselective adsorption is preserved with increasing cluster size, for tetramers and hexamers, however, DFT calculations show that the opposite surface-bound conformer is more stable, as illustrated in Figures 5 with the tetramer unit composed of 4 hydrogen-bonded $R_C$-$S_O$-2-BuOH molecules. Furthermore, tetramers composed of four hydrogen-bonded $R_C$-$S_O$-2-BuOH molecules have been shown to be 35 meV more stable than a tetramer consisting of molecules with the opposite surface-bound $R_C$-$R_O$ enantiomer (SI figure S1g). Perhaps, the decrease in the BuOH-metal interaction for this cluster size, as indicated in figure 4, diminishes the geometrical strain imposed on the bond between the oxygen lone pair and the surface, allowing for the $R_C$-$S_O$-2-BuOH conformer to be more energetically favorable at this cluster size.

The interesting ramification of this result is that the intrinsic chirality of gas phase 2-BuOH directs the chirality of the second chiral center of the adsorbed state of the molecule, resulting in 2-BuOH hydrogen-bonded tetramers having the same chirality at the surface-bound oxygen atom. As a consequence, at high surface coverage, extended zig-zag hydrogen-bonded chain structures that form from the interaction between opposite surface-bound enantiomers, as illustrated in Figure 5, and observed for methanol at a comparable coverage, will be energetically disfavored for 2-BuOH since these would require half of the 2-BuOH molecules to adsorb in an unfavorable orientation at the oxygen atom. This is consistent with experimental observations, at the monolayer regime, of homochiral tetrameric units and not of heterochiral zig-zag chains.

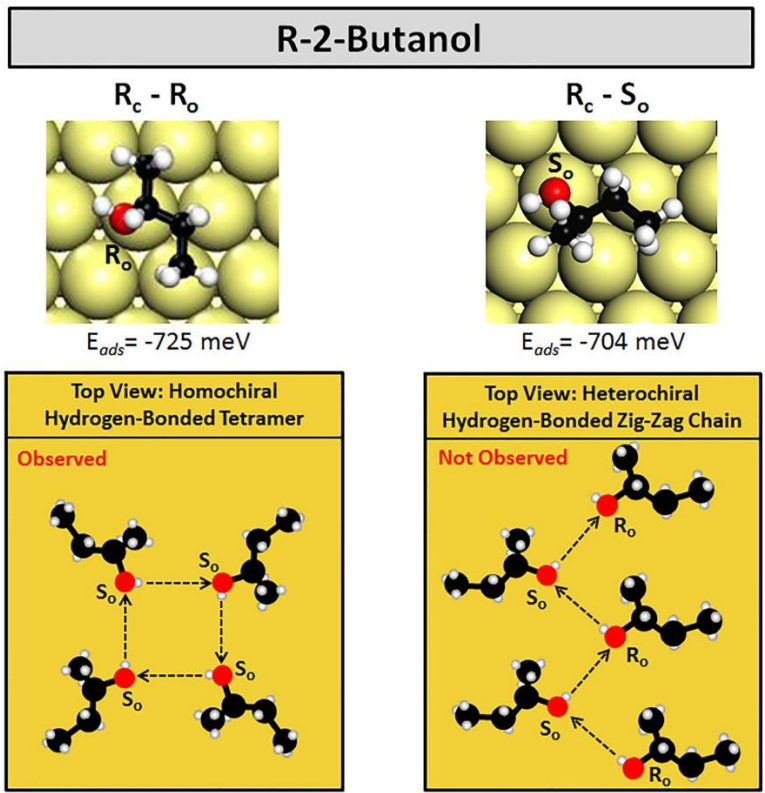

**Figure 5**. Top panels illustrate that when R-2-BuOH monomers bind to the surface through an oxygen lone pair a second chiral center ($R_o$ or $S_o$) is created. These molecules, which have the same intrinsic chirality ($R_c$), are considered surface-bound diastereomers. The lower panel shows a top view schematic of the experimentally observed hydrogen-bonded tetramer consisting of four 2-BuOH molecules with the same intrinsic chirality ($R_c$) and the same surface-bound chirality at the oxygen atoms ($S_o$) and a top view schematic of a chain structure, which is not experimentally observed at the monolayer coverage, which requires both surface-bound diastereomers of R-2-BuOH to be present on the surface.

A related observation was reported for S- and R-3-pyrroline-2-caboxylic acid (PCA) on a Cu(110) surface.[47-48] STM and DFT data showed that the intrinsic chirality and certain structural features of PCA leads to the adsorption of one conformer, creating a second chiral center upon adsorption to the substrate. The Cu(110) surface was rendered chiral as the intrinsic chirality in the admolecule was preserved and transferred to the surface bond which further promoted the assembly of homochiral overlayers on the surface.[47-48]

**B. Build up of long-range chiral networks:** Having established what the basic building block in 2-BuOH adsorption is we now look at the extended structure of the overlayer, in particular the domain structure observed. If we look again at Figures 1c and 2a these show examples of two 2-BuOH tetramers of the same chirality but with two different rotational orientations, with respect to the surface lattice. Owing to the six-fold symmetry of Au(111) and the slight rotation of the units from the high symmetry direction of the surface, tetramers with three different rotations would be expected and indeed were observed experimentally. High-resolution STM images of the three rotational domains that make up the R-2-BuOH film show that at the 1 ML regime the first rotational island is oriented +5° (clockwise) from the close-packed, or the $[1\bar{1}0]$, symmetry direction of the underlying Au surface (Figure 6a-c), rendering the network an extended array of chiral "pockets". The other two rotations are oriented +120° and +240° from rotation 1, respectively. We never observe the opposite rotations for R-2-BuOH which provides further evidence that: (1) the intrinsic molecular chirality transfers to the second chiral center of the adsorbed state at the oxygen atom and (2) the enantiospecific interaction of the chiral O-H entities lead to the formation of homochiral tetramers with two complementary chiral centers that are rotated ~5° from the high symmetry directions of the surface lattice. This is supported by our experiments which show that adsorption and assembly of the opposite enantiomer, S-2-BuOH, leads to the same structures with opposite rotational (i.e. -5°, -65°, -125° from the $[1\bar{1}0]$ direction) and mirror symmetries.

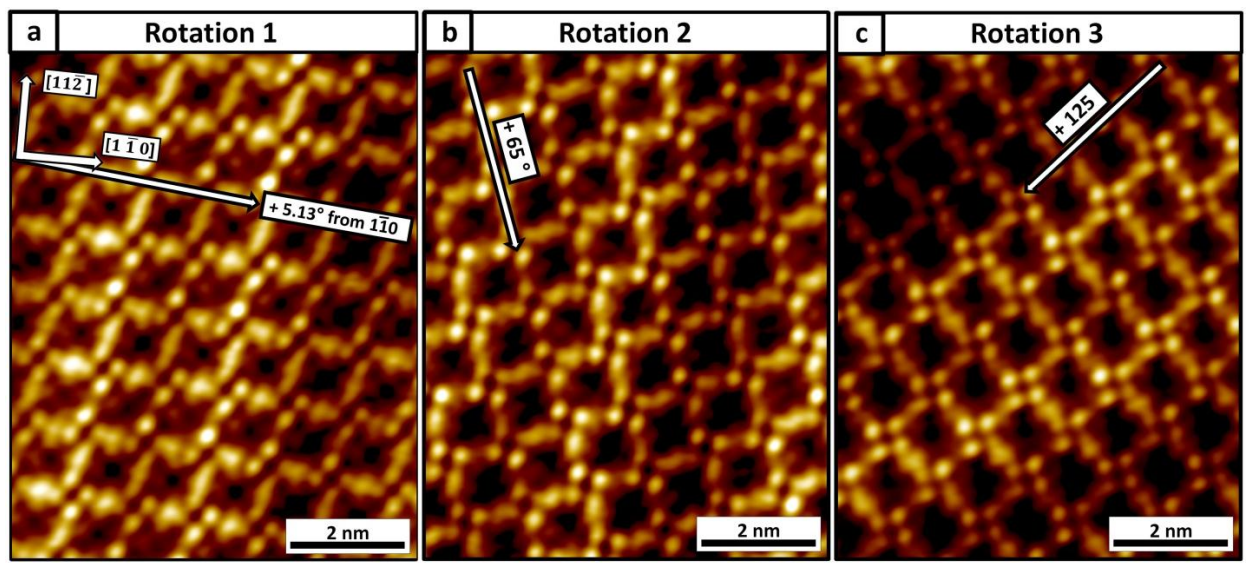

**Figure 6.** High-resolution STM images of R-2-BuOH at 1 ML coverage on a Au(111) surface. Three rotational domains were observed and are labeled Rotation 1-3. (a) Rotation 1 is oriented +5° (clockwise) from the close-packed, or the $[1\bar{1}0]$, symmetry direction of the underlying Au(111) surface, as indicated. Rotations 2 and 3 are rotated +65° and +125°, respectively, from the $[1\bar{1}0]$ direction.

To test this proposed model we also imaged defects (which were rather uncommon) and island edges of the chiral networks that were present at the submonolayer regime. At < 1 ML coverage, the dominant features on the surface are isolated islands composed of the tetramer base units that have adopted one of the three possible rotational domains as depicted in Figure 7 a-c. The bright pair-wise soliton walls from the herringbone reconstruction are visible under the 2-BuOH networks in the topographic image. Our model, which indicates that self-assembly is based upon the formation of hydrogen bonded tetramers that are linked to each other via vdW interactions, is supported by the striking symmetry of the islands which have straight edges that lie along low symmetry directions of the surface lattice. We also find that all the edges and corners of the network islands are terminated by an intact 2-BuOH tetramer, as seen in Figure 7d and 7e. Furthermore, in rare cases, a single core tetrameric structure is incomplete and its effect on the square network within a domain or at the edge of an island is evident as it causes a disruption in the square pattern, as shown in Figures 7e-g. These observations support the conclusion that enantioselective formation of the hydrogen-bonded 2-BuOH tetramers and their self-assembly via vdW interactions with matching rotational symmetry drive a second level of ordering to yield the long range self-assembly of this simple chiral alcohol.

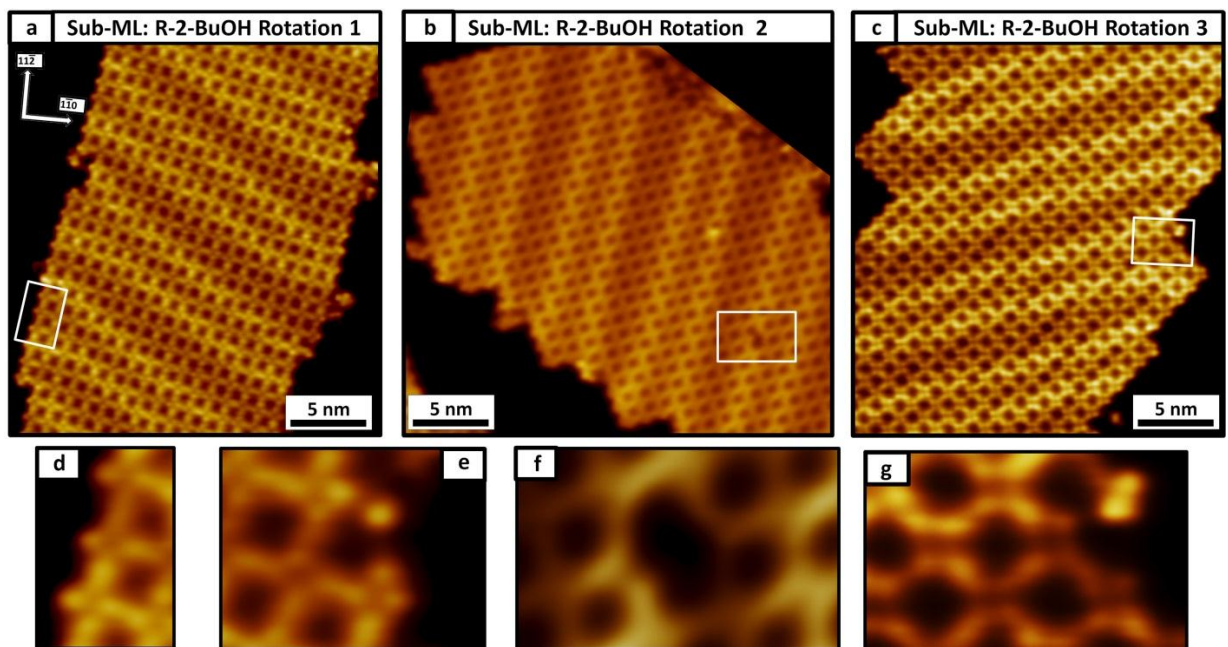

**Figure 7.** Zoomed-in, high-resolution STM images of R-2-BuOH rotational domains at submonolayer coverage. (a, b, c) Three rotational domains were observed. As in the ML regime, domains consist of an array of square pockets, domains are rotated +120° relative to each other, and their orientations are not aligned with the high symmetry directions of the underlying Au(111) surface. (d) Zoomed- in image of the edge of the domain labeled rotation 1, showing that the edges are terminated by an intact hydrogen-bonded tetramer. (e, f, g) High resolution STM images showing that on rare occasions, in the absence of an intact hydrogen-bonded tetramer the square pattern is disrupted at domain edges and within the domain.

Figures 8a and 8e show large-scale STM images of R- and S-2-BuOH, respectively, at full monolayer coverage. The larger STM images reveal that the adlayer is composed of the three rotational domains that come together to form a continuous film over the Au surface. Faint depressed lines visible in the overlayer correspond to rotational domain boundaries between different rotationally oriented islands. Figures 8 b-d show zoomed-in images of the three rotational domains observed for the 1 ML R-2-BuOH system. Even at this high molecular coverage, the herringbone reconstruction of the underlying Au surface is still present. In order to investigate the interaction of 2-BuOH with the surface itself we measured the spacing of the herringbone reconstruction which has shown to be a useful guide for the interaction strength of weakly adsorbed molecules.[22-25, 54-56] On the clean Au surface, the herringbone spacing is 6.3 nm.[41] Measurements of the herringbones visible under the 2-BuOH networks indicate that they have a spacing of 6.4 ± 0.4 nm, which shows that the herringbone reconstruction remains unperturbed. These measurements indicate that 2-BuOH molecules do not bind to the Au surface as strongly as species like thioethers, which eject Au atoms from the surface causing a perturbation and partial lifting of the herringbone reconstruction.[22-25, 54-56] Since in the 2-BuOH system, the herringbone reconstruction remains unperturbed the isolated brighter protrusions observed on the networks (Figure 8) are attributed to 2-BuOH molecules in the second layer as opposed to ejected Au atoms.

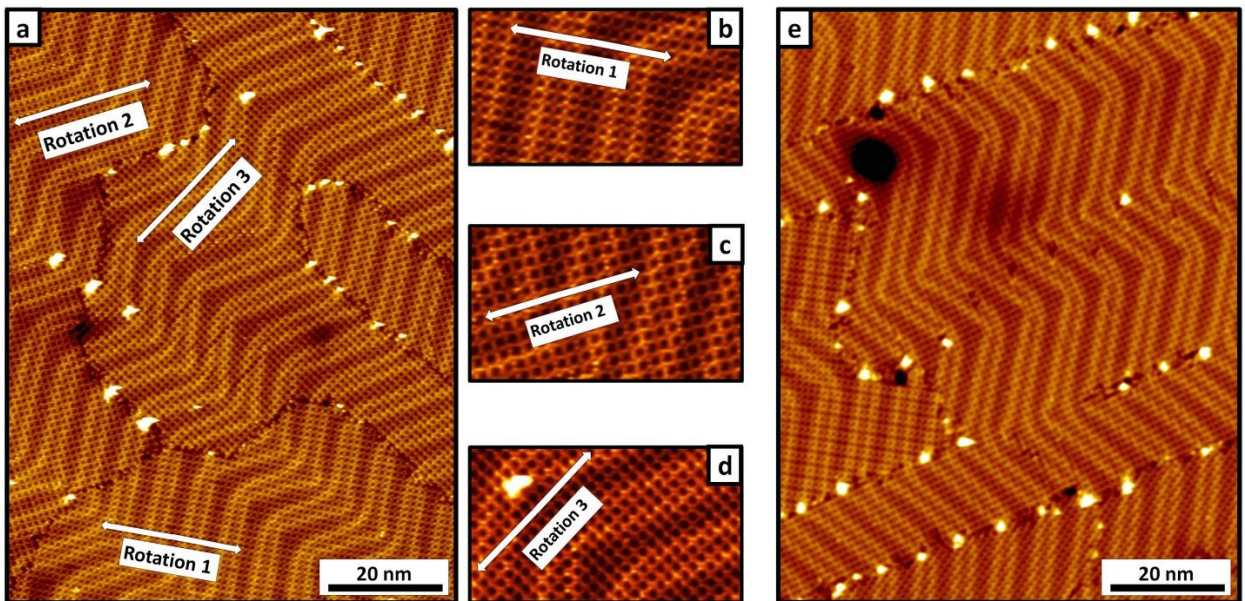

**Figure 8.** Large-scale STM images of 2-BuOH chiral networks (a) 1 ML R-2-BuOH on Au(111). Three rotational domains come together to form a thin film on the Au surface. The observed fissure lines are domain boundaries between different rotational islands. The herringbone reconstruction is still present with the spacing between neighboring soliton walls measuring 6.4 ± 0.4 nm, indicating that the Au surface reconstruction remains unperturbed. The bright protrusions on the R-2-BuOH network are attributed to second layer R-2-BuOH molecules. (b, c, d) Zoomed-in images of the three rotational domains present in (a). (e) S-2-BuOH film on a Au(111) surface at the 1 ML regime. The rotational domains in the S-2-BuOH system are mirror images of those in (a).

Each unit of the chiral domains is made up of four tetramers arranged in a square, as illustrated by the molecules superimposed over the network in Figure 9(a). It seems reasonable to assume that long-range vdW forces are at least partially responsible for holding the tetramers together in an extended overlayer structure. Indeed DFT calculations for gas phase films of tetramers (in the structures they adopt in the adsorption system) do indeed show that vdW forces create an attractive interaction between the tetramers. Figure S3 in the SI section shows that the ideal 2-BuOH tetramer distance calculated by DFT, in the gas phase, has an interaction energy equal to ~-50 meV, which is larger than many dipole-dipole interaction energies for more dipolar molecules, further supporting that van der Waals interactions makes a significant contribution towards the attractive interactions between tetramers and drive the supra-molecular self-assembly on this surface.[57]

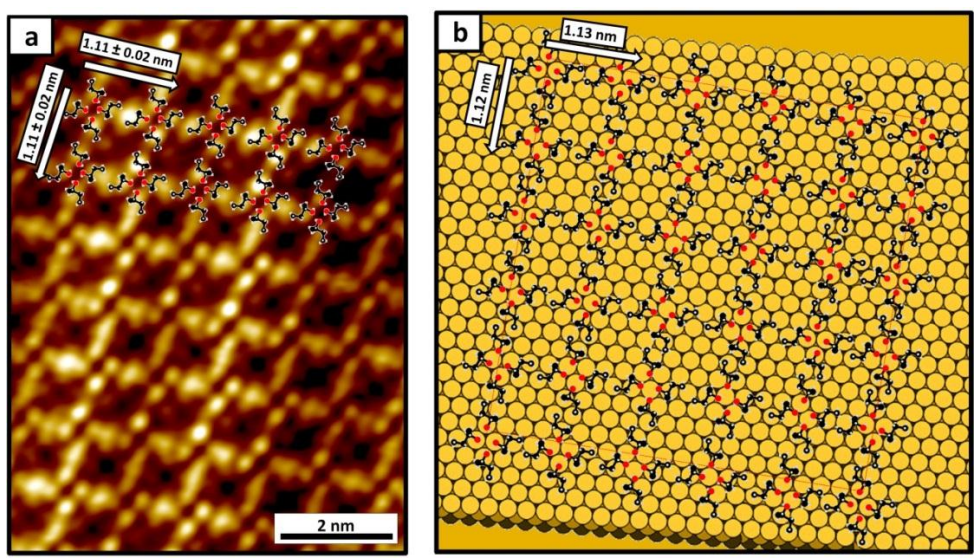

**Figure 9.** (a) High-resolution STM image of an R-2-BuOH domain (rotation 1). Schematic showing the assembly of R-2-BuOH hydrogen-bonded tetramers to form square chiral units is superimposed over the STM image. Experimental measurements for the square chiral pockets are indicated. (b) Proposed unit cell for the 2-BuOH chiral domain consisting of hydrogen-bonded tetramers that assemble via van der Waals interactions between alkyl tails to form an array of square chiral pockets on an unreconstructed Au(111) surface.

In terms of the overall network structure, the proposed unit cell is shown in Figure 9(b); each square "pocket" in the network is based on the vdW interaction between alkyl tails at the exterior of the four hydrogen bonded, like-rotated tetramers. These pockets measure 1.11 ± 0.02 nm x 1.11 ± 0.02 nm. One can think of the network as being composed of hydrogen bonded *hydrophilic* cores (the tetramers) assembled into a square network by vdW attraction between alkyl tails of neighboring tetramers leading to the formation of a regular array of *hydrophobic* chiral pockets in the network. The schematic shown in Figure 9(b) consists of 5 x 5 square chiral pockets with dimensions of 1.13 x 1.12 nm, which agrees well with experimental measurements.

## 4. CONCLUSIONS

In summary, STM and DFT were used to show that the intrinsic chirality in a simple chiral alcohol, 2-BuOH, can be transferred to the adsorbed state of the molecule to yield homochiral assemblies at a variety of length scales that exhibit different levels of ordering ranging from a local, lower level of order driven by hydrogen-bonding to a higher degree of long range order dominated by van der Waals interactions. Upon adsorption, 2-BuOH monomers develop a second chiral center at the surface-bound oxygen and DFT calculations predict that the formation of just one surface-bound enantiomer is energetically preferred. In other words, the intrinsic chirality of the 2-BuOH molecule dictates the binding geometry of the adsorbed species, resulting in 2-BuOH molecules having the same surface-bound chirality in a given system. DFT calculations confirmed the high stability of surface bound tetramers consisting of four 2-BuOH molecules involved in a hydrogen-bonded network, with each molecule donating and receiving a hydroxyl proton. The 2-BuOH tetramer adopts a planar configuration on the Au surface, with each oxygen atom binding on nearly atop sites. Moreover, hydrogen-bonded tetramers exist in only three rotational orientations, which is consistent with the symmetry of the underlying Au substrate.  Due to this enantiospecific adsorption, 2-BuOH tetramers are homochiral hydrogen-bonded networks. At a larger scale, these homochiral hydrogen-bonded tetramers with the same rotational orientation assemble via vdW interactions between alkyl tails to form large ordered domains that consist of a network of square chiral "pockets" that exist in three different domain orientations. Furthermore, the orientations of these ordered domains do not align with the high symmetry directions of the underlying Au lattice, rendering the whole Au surface chiral at monolayer 2-BuOH coverage. Future experiments will focus on probing the enantioselective properties of the 2-BuOH chiral "pockets" to determine, for instance, if they are able to mediate enantiospecific adsorption of chiral probe molecules or capable of facilitating asymmetric reactions when grown on catalytically active alloyed Au surfaces.

**Funding Sources**

This work was supported by the Department of Energy BES under grant DE-FG-85ER20203. ML thanks the NSF for a Graduate Research Fellowship. JC is supported by the MINECO through a Ramón y Cajal Fellowship and acknowledges support by the Marie Curie Career Integration Grant FP7-PEOPLE-2011-CIG: Project NanoWGS and The Royal Society through the Newton Alumnus scheme. CS thanks the Dreyfus Foundation for a Teacher-Scholar award. A.M. is supported by the European Research Council under the European Union's Seventh Framework Programme (FP/2007-2013) / ERC Grant Agreement number 616121 (HeteroIce project). A.M is also supported by the Royal Society through a Royal Society Wolfson Research Merit Award. We are also grateful for computational resources to the London Centre for Nanotechnology, UCL Research Computing, and to the UKCP consortium (EP/F036884/1) for access to Archer.

**Notes**

The authors declare no competing financial interest.